# Controlling supercurrents and their spatial distribution in ferromagnets

Kaveh Lahabi, Morten Amundsen, Jabir Ali Ouassou, Ewout Beukers, Menno Pleijster, Jacob Linder, Paul Alkemade, and Jan Aarts*

**Spin-triplet Cooper pairs induced in ferromagnets form the centrepiece of the emerging field of superconducting spintronics [1,2]. Usually the focus of research is on the spin polarization of the triplets, potentially enabling low-dissipation magnetization switching and domain wall motion. However, the fundamental mechanism for generating triplet pairs [3,4] also permits control over a parameter which has not been addressed before, namely the spatial distribution of the supercurrent. Here we demonstrate this control by tailoring distinct supercurrent pathways in the ferromagnetic weak link of a Josephson junction. Combining micromagnetic simulations with three-dimensional critical current calculations, based on the Usadel description of mesoscopic superconductivity [5], we designed a disk-shaped structure with a magnetic vortex, which induces *two* distinct supercurrent channels across the junction. The design was successfully tested with superconducting quantum interferometry (SQI). Moreover, we show how the position of the pathways can be controlled by moving the vortex with a magnetic field. This novel approach allows adaptable supercurrent paths to be dynamically reconfigured to switch between different functionalities in the same device.**

Converting spin-singlet Cooper pairs to equal-spin triplets requires carefully designed interfaces between a conventional superconductor (S) and a ferromagnet (F). The process requires both spin-mixing and spin-rotation, and can be brought about by magnetic inhomogeneities at the interface [3]. One method to realize this is to place a thin ferromagnet F' at the S/F interface, and make the magnetization of F and F' non-collinear [4]. This method was recently implemented in Josephson junctions described by 1D geometries, where the supercurrent amplitude was controlled by varying degrees of magnetic non-collinearity (MNC) [6,7,8]. The present letter establishes a different direction, where the focus is not the supercurrent amplitude. Instead, the central goal is to exert dynamic control over the triplet generator which in turn determines where the supercurrent spatially flows. We demonstrate for the first time that the position and number of supercurrent channels can be altered in a dynamic fashion. Such behaviour is intrinsically higher-dimensional and can pave the way for novel hybrid devices in superconducting electronics.

The device consists of a disk-shaped planar Josephson junction involving a multilayer of Co/Cu/Ni/Nb, as shown in **Fig. 1a**. A central trench cuts the top superconducting Cu/Ni/Nb layers in two halves, here connected via a Co weak link. The disk design combines two crucial elements. First, the magnetic moments in Co are arranged in plane and orthogonal to the trench between the superconducting electrodes, while the moments in Ni lie also in plane but parallel to the trench. Micromagnetic simulations show that this geometry results in a well-defined magnetic ground state with a high degree of MNC, a condition optimal for generating triplets (**Fig. 1c-e**). An equally important element is that the disk shape creates a magnetic vortex state in the Co. This vortex produces a distinct suppression of MNC at the centre of the disk (**Fig. 1e**), which will be used to distribute the supercurrent in Co over two channels. The MNC suppression is due to the local out-of-plane magnetization at the vortex core, which turns the magnetic moments in the Ni also out-of-plane and, hence, collinear to the Co moments. Incidentally, the *in-plane* exchange field gradient of a magnetic vortex, without a second ferromagnet, has also been proposed to generate long-ranged triplets [9,10].

To investigate whether a supercurrent can be expected, we numerically simulate the critical current density passing through the Josephson junction by solving the quasiclassical Usadel equation [5] in 3D using the magnetization texture obtained from the micromagnetic simulations. We do this by means of the finite element method, using the finite element library libMesh [11] in a similar fashion as in Ref. [12]. The superconductors are modelled as bulk, with a phase difference of $\Delta\phi = \frac{\pi}{2}$. In **Fig. 2a** the discretized model is shown. To reduce the calculation time we truncated the otherwise circular geometry to a width of 40% of the disk diameter, as the currents farther away from the trench are negligible. The results are shown in **Fig. 2b**, where it can be seen that the critical current is suppressed at the centre of the disk, thereby effectively creating two separate current channels.

Turning to the results, our junctions show zero resistance and finite critical currents $I_c$ below 3 K (**Fig. 3a-b**). We examine their supercurrent profile by applying an out-of-plane magnetic field $B_z$ resulting in a current interference pattern. As demonstrated by Dynes and Fulton [13], the shape of such a superconducting quantum interference (SQI) pattern is given by the Fourier transform of the position-dependent critical current density across a junction $J_c(x)$ through

$$I_c(B_z) = \left| \int_{-R}^{R} dx\, J_c(x)\, e^{2\pi i L B_z x/\Phi_0} \right|$$

where $L$ is the effective length of the junction, $R$ is its lateral width (here the disk radius), and $\Phi_0 = h/2e$ is the superconducting flux quantum. In a typical junction, the uniform distribution of supercurrent density $J_c$ ($J_c(x)$ = constant) leads to the well-known Fraunhofer interference pattern with a sinusoidal current-phase relation given by $I_c(B_z)/I_c(0) \sim |\sin(\pi\Phi/\Phi_0)/(\pi\Phi/\Phi_0)|$. Characteristic for the Fraunhofer pattern is a central lobe that is twice as wide as the side lobes (as in **Fig. 3e**). These oscillations decay with a $1/B$ dependence. Different device configurations may introduce deviations from the standard pattern, but the described relative widths of the lobes persist as a common feature in all Josephson junctions, since it represents a single-slit interference pattern. In contrast, we expect our disk to exhibit a double-slit interference pattern. This is characterized by slowly decaying sinusoidal oscillations with $\Phi_0$-periodicity, where all lobes have the same width (**Fig. 3c,d**). These patterns are typical for superconducting quantum interference devices (SQUIDs) which, contrary to our device, consist of two *individual* junctions operated in parallel.

As shown in **Fig. 3c-d**, the period of the oscillations in our disk device is $7.8$ mT (i.e. fluxoid quantization over an effective area of $2.65 \times 10^{-13} \mathrm{m}^2$), and appears to be temperature-independent. This behavior is typical for planar junctions with incomplete screening by the thin superconducting electrodes. In this case, the effective screening length is not determined by the London penetration depth, $\lambda_L$, but rather by the geometrical boundaries of the device [14].

Qualitatively, the $I_c(B_z)$ data in **Fig. 3c-d** already foretell the presence of two supercurrent channels: the width of the central lobe is comparable to that of the side lobes, and the oscillations decay far more gradually in field than as $1/B$. Two-channel interference patterns were recently observed in junctions with topological weak links [15,16,17], where the two-slit interference is a result of edge-dominated transport caused by band bending. In our junction this is due to the suppression of triplet supercurrent by the (controllable) magnetic vortex core.

To illustrate the contrast with single-slit interference in a similar device configuration, we prepared a disk junction without the Ni layer and retaining a thin layer of Cu/Nb at the bottom of the trench. This provides a non-magnetic path in the weak link, allowing singlet correlations to contribute to junction transport. Indeed, we observe a typical

Fraunhofer-like interference pattern with a two times wider central lobe, shown in **Fig. 3e**.

**Fig. 3f** shows the supercurrent density profiles extracted from Fourier analysis of the measured interference patterns. A full description of this method can be found in the Supplementary. Two distinct transport channels are clearly visible in all extracted profiles. Comparing these results with the simulations, the screening supercurrents appear to follow narrower paths, located near the centre of the disk. This may be attributed to the varying length of the disk-shaped electrodes. The original method in Ref [13] assumes the effective magnetic length of the junction to be uniform and equal to $L = d + 2\lambda_L$, with $d$ the gap between the electrodes, while screening in our junctions is limited by the geometrical boundaries rather than by $\lambda_L$. As the size of the disk-shaped electrodes decreases towards the edges of the junction ($x \to \pm R$), screening becomes less effective, and the supercurrent density diminishes. Note that the Fourier analysis yields the distribution of the *screening* supercurrent (driven by an applied flux), while the simulated supercurrent distribution in **Fig. 2** describes current-biased transport in the *absence* of external magnetic fields. Hence, they provide different but complementary information about electrical transport in our system.

Having established the principal role of MNC in shaping the supercurrent, we also examine the possibility of controlling them by altering the MNC profiles using an in-plane field $B_y$. **Fig. 4a** shows the measured currents $I_c(B_y)$ together with the micromagnetic MNC calculations for the various magnetic states. In the first regime (i), we alter the MNC profile by bringing the vortex core to one side of the disk. In (ii), we remove the vortex, thereby suppressing the supercurrent. This suppression is due to the antiparallel configuration of the ferromagnets, occurring through the increase of stray fields from the Co layer after removal of the vortex. In the third regime (iii) the Ni is magnetized parallel to the Co. First, this recovers $I_c$ as MNC re-emerges over the entire disk. Next, the magnetization in both layers becomes parallel and $I_c$ disappears again. **Fig.4b**. shows the results of a field sweep from a high positive field to a negative field and back. We observe a clear hysteresis, and striking variations in $I_c$. The distinct features of this pattern correspond to the variation in MNC as we switch between different magnetic states (including i-iii). More details are presented in the supplementary.

Spin-triplet supercurrents in ferromagnets have been bearing the promise of dissipationless use of spin-polarized currents. This study opens up a completely different direction, in which the focus is not the homogeneous amplitude of the supercurrent, but rather the dynamical control over its spatial distribution. This can lead to novel hybrid devices for superconducting electronics. Moreover, our extensive use of simulations, both of the micromagnetic configurations and of the supercurrents themselves, allow for detailed design and understanding before the actual fabrication of the hybrid device. The next step will be to introduce magnetization dynamics. Magnetic vortices or domain walls can be moved with pulses in the GHz regime, and this can also be simulated. Directing supercurrents then becomes possible on nanosecond timescales, opening the way for high-speed superconducting electronics.

## Acknowledgements

The authors would like to thank S. Goswami, A. Singh, M. Kupryianov, S. Bakurskiy and J. Jobst for valuable discussions and comments. This work was supported by the Netherlands Organisation for Scientific Research (NWO/OCW), as part of the Frontiers of Nanoscience program.

## Author Information

### Affiliations

Huygens – Kamerlingh Onnes Laboratory, Leiden Institute of Physics, University Leiden. Leiden, The Netherlands.
Kaveh Lahabi, Ewoud Beukers, Menno Pleijster, Jan Aarts

Department of Physics, Norwegian University of Science and Technology, N-7491 Trondheim, Norway.
Morten Amundsen, Jabir Ali Ouassou, Jacob Linder

Kavli Institute of Nanoscience, Delft University of Technology, Delft, The Netherlands

**Contributions**

K.L and J.A conceived the disk geometry, K.L and E.B performed the micromagnetic simulations, M.A, J.A.O. and J.L. performed the supercurrent simulations and assisted in the Fourier analysis, K.L, M.P and P.F.A fabricated the devices, K.L and M.P performed the measurements. All authors contributed to discussions.

**Competing financial interests**

The authors declare no competing financial interests.

**Corresponding author**

Correspondence to: Jan Aarts (aarts@physics.leidenuniv.nl)

# Methods

**Device fabrication.** Multilayers of Co(60nm)/Cu(5nm)/Ni(1.5nm)/Nb(45nm) were deposited on unheated $SiO_2$-coated $Si$ substrates by $Ar$ sputtering in an ultra-high vacuum chamber (base pressure below $10^{-8}$ Pa). The thickness of the Co and the diameter of the disk are chosen to ensure stabilization of a magnetic vortex [18,19]. The 5 nm Cu layer is used to avoid exchange coupling between the layers. The thickness of the Ni layer was tuned for optimal triplet generation in similar systems [20,21]. The samples were subsequently coated with Pt (7nm) to protect them from oxidation and to reduce the damage introduced by $Ga^+$ ions during focused ion beam (FIB) processing.

A combination of electron-beam lithography and FIB milling ($50$ pA $Ga^+$ beam current) was used to structure the disks. Next, FIB with $1$ pA current was applied to open the sub-20 nm gap that forms the junction. The trench depth is controlled by the duration of milling. The $1$ pA beam current provided sufficient timespan (several seconds) to vary the depth in a controlled manner. The device used for investigating single-slit transport was subject to the same processing steps, with the following exceptions. First, the multilayer was deposited without Ni to minimize triplet generation. Second, when creating the weak link, the duration of FIB milling was reduced by 20% to retain a layer of Cu / Nb at the bottom of the trench. This provides a non-magnetic path for singlet supercurrent in the weak link (on top of the Co).

The trench is presumably deeper near the sides of the disks (where sputtered atoms can escape more easily) than at the centre. Hence, in contrast to triplets, singlet correlations would favour the centre of the disk where a nonmagnetic channel may be still present on top of the Co.

**Measurements.** The magnetic properties of Co and Ni used in our devices were characterized by ferromagnetic resonance experiments and SQUID magnetometry. Transport measurements were performed in a Quantum Design Physical Properties Measurement System where samples could be cooled down to $2.1$ K. The current-voltage characteristics were carried out in a four-probe configuration using a current-biased circuit and a nanovoltmeter. The critical current was determined using a voltage criterion: $V > 0.3\ \mu$V for SQI and $V > 0.1\ \mu$V for the measurements with an in-plane field.

The virgin state was measured directly after fabrication (see the supplementary). Prior to the $I_c(B_z)$ measurements presented in the letter, the magnetic state of the sample was conditioned by applying a $2.5$ T out-of-plane field at $10$ K. The sample was stored in a UHV chamber for 106 days and re-wired to a different puck, and the same measurements were repeated using a different magnet. We were able to reproduce the same $I_c$ patterns, and no discernable changes in transport characteristics (e.g. $\text{R(T)}$ or $I_c$) were observed.

**Micromagnetic simulations.** The finite element micromagnetic calculations were carried out using the Object Oriented Micromagnetic Framework (OOMMF) [22] The multilayer is divided into a three-dimensional mesh of $5\text{nm}$ cubic cells. The exchange coefficient and saturation magnetization of Co were set to $30 \times 10^{-12}$ Jm$^{-1}$ and $1.40 \times 10^6$ Am$^{-1}$, respectively, while for Ni these values were $9.0 \times 10^{-12}$ Jm$^{-1}$ and $4.90 \times 10^5$ Am$^{-1}$. The Gilbert damping constant $\alpha$ was set to $0.5$ to allow for rapid convergence. The direction of anisotropy was defined by a random vector field to represent the polycrystalline nature of the sputtered films. The Usadel calculations are based on static micromagnetic simulations of a multilayer disk with a diameter of $1\mu\text{m}$. For simulations with an applied in-plane field (Fig.4), the disk design was extended to include the leads used for transport measurements in the actual device (see Supplementary Fig. S3).

## Captions

Fig.1 | **Micromagnetic simulations and the device layout.** **a.** Schematic of the device layout. **b.** False-colour scanning electron microscope image of a device. The disk is structured with $\text{Ga}^+$ focused ion beam (FIB) milling. The junction is formed by opening up a gap in the top Nb/Ni/Cu layers, leaving only Co in the weak link (see Methods for more details). **c.** plane view of the magnetic states of Co and Ni layers in the disk (from 3-D OOMMF simulations). The pixel colour scheme, red-white-blue, scales with the magnetization along $y$. Magnetic moments in Ni tend to align with the gap which defines the junction, while the vortex configuration in Co arranges the magnetic moments perpendicular to it. This provides a high degree of MNC for triplet generation. The curled magnetic structure of the vortex is also highly effective in minimizing the stray fields from Co, which otherwise would dominate the Ni

magnetization, compromising our control of MNC. **d.** represents our method to obtain the MNC profile. For each cell at the top of the Co layer, we determine the angle $\vartheta$ between its magnetization vector and that of the Ni cell above. **e.** spatially resolved MNC profile calculated from the simulation results of **c**. The observed suppression of MNC (blue region) at the centre of the junction is a result of interlayer dipole coupling at the vortex core.

Fig.2 | **Numerical simulation of the critical current. a,** the discretized model (or mesh) used in the numerical simulation of the critical current. Since the triplet current is mostly concentrated in the immediate vicinity of the trench, the mesh density, and hence the accuracy, is higher in this region. For the same reason, the regions farthest away from the trench have been removed to reduce the calculation time. **b,** the critical current density divided by a factor $J_0 = \frac{N_0 eD\Delta}{2\xi}$, where $N_0$ is the density of states at the Fermi level, $D$ is the diffusion constant, $\Delta$ is the superconducting gap and $\xi$ is the superconducting coherence length. For clarity, currents lower than $10^{-7}J_0$ are not shown. The lower figure shows a slice through the centre of the trench where it is seen how the current passes through the trench in two separate channels on either side of the centre of the disk.

Fig.3 | **Junction transport and SQI patterns. a** and **b**, show the temperature-dependence of resistance and I-V traces respectively. R(T) has two distinct transitions corresponding to the superconducting electrodes (at 5.5 K) and the junction (below 3.2K). **c** and **d**, SQI patterns taken at 2.1 K and 2.8 K respectively. The patterns show clear double-slit (SQUID-like) interference, with all lobes having the same width. **e,** single-slit interference pattern from a disk where transport is dominated by singlet correlations via a non-magnetic medium in the weak link. **f,** the current density profiles constructed from the Fourier analysis of the SQI patterns at 2.1 K , 2.5 K and 2.8 K.

Fig.4 | $I_c(B)$ **and the corresponding magnetic states from micromagnetic simulations. a,** measured $I_c$ values and the corresponding MNC profiles, as $B_y$ magnetizes the measured the system. **(i)**, the vortex core moves along the junction (perpendicular to field direction) to the side of the disk. Highly MNC regions are continuously present and appear to follow the position of the vortex core. **(ii)** The vortex configuration, which has been effective in suppressing the stray fields, vanishes as Co continues to magnetize along $+y$. This leads to a negative dipole field from Co which dominates the effective field acting on Ni. As a result, Ni gets magnetized antiparallel to Co (along $-y$), hence the suppression of MNC and $I_c$. **(iii)**, the increasing applied field begins to compensate the contribution of Co stray fields,

ultimately reversing the Ni magnetization. The change in the magnetic orientation associated with the reversal leads to a distinct (re-)emergence of MNC that gradually fades away above $60\,\text{mT}$ - as Ni magnetization aligns with Co. **b,** $I_c(B_y)$ measured during field reversal.

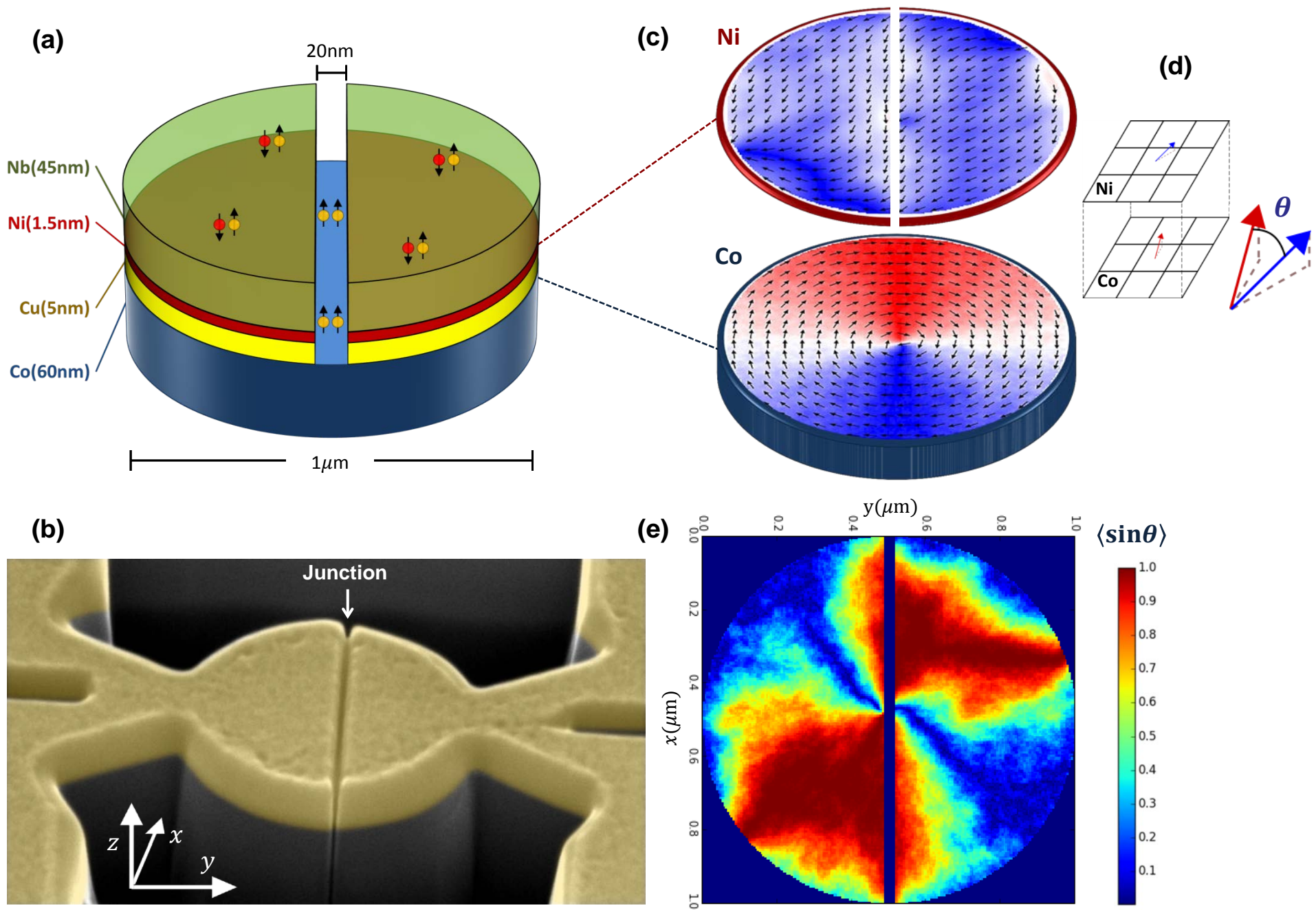

**Figure 2**

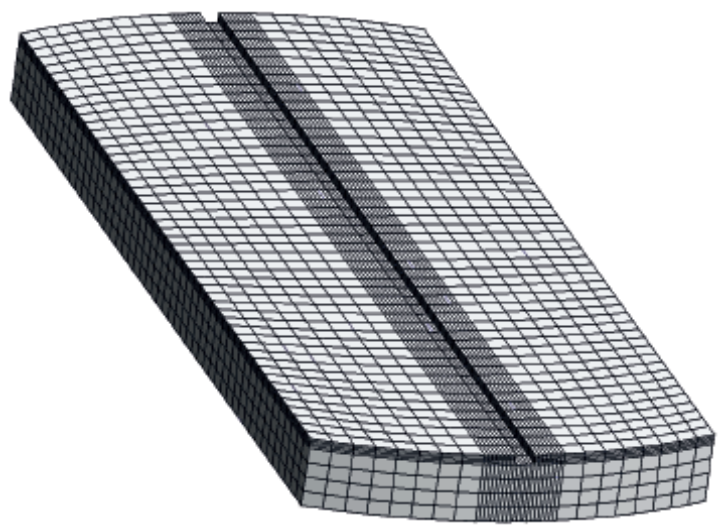
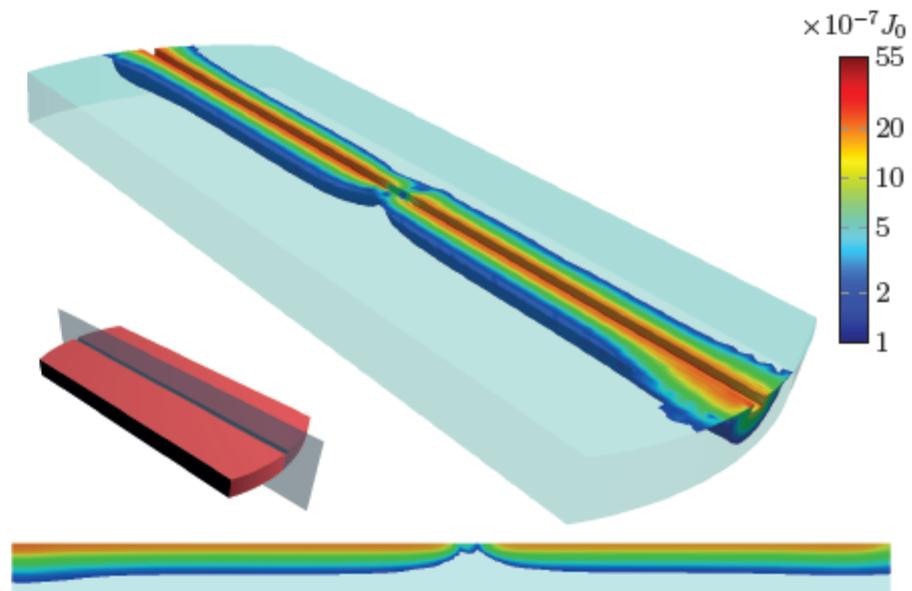

(a)  (b)

**Figure 3**

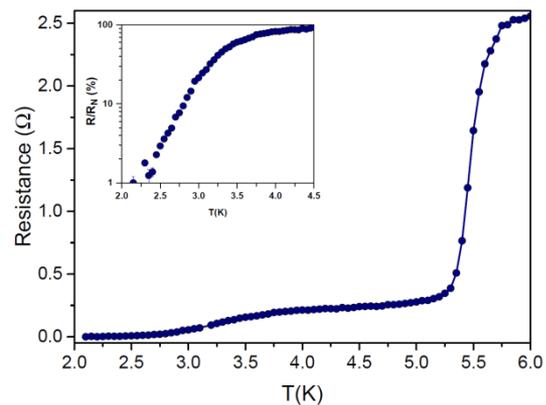
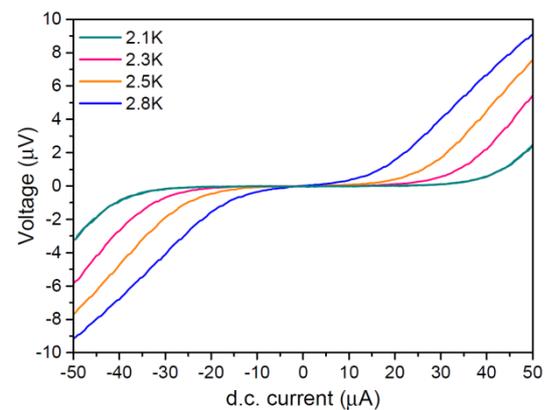
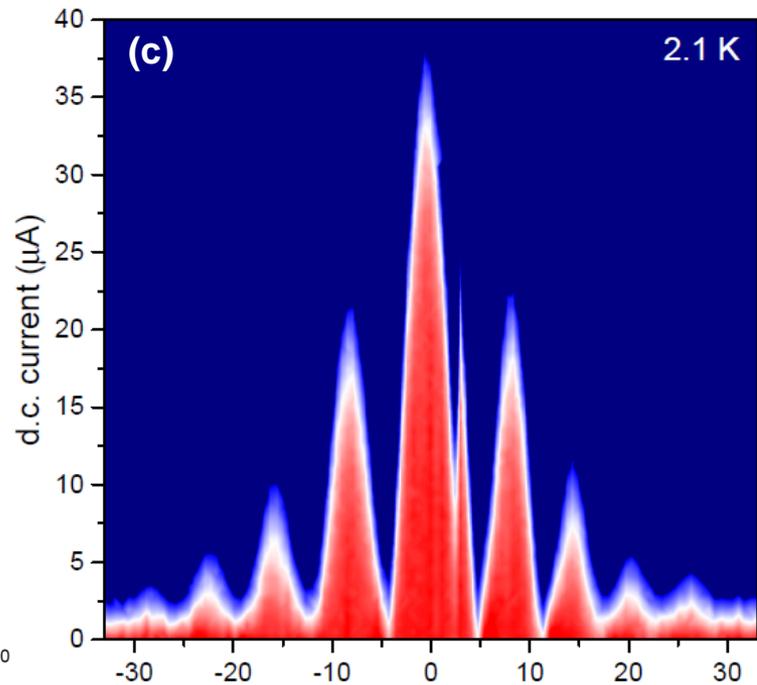
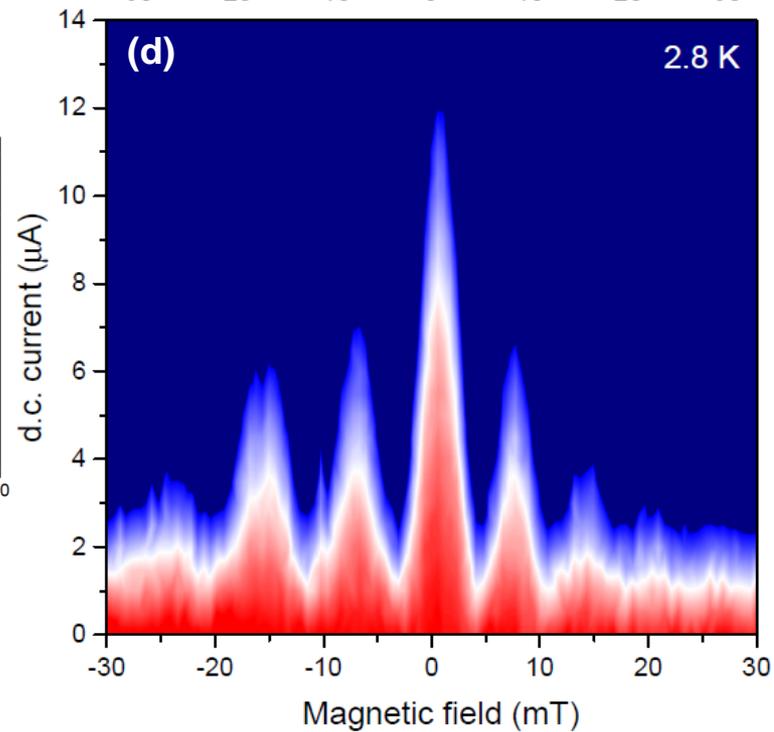
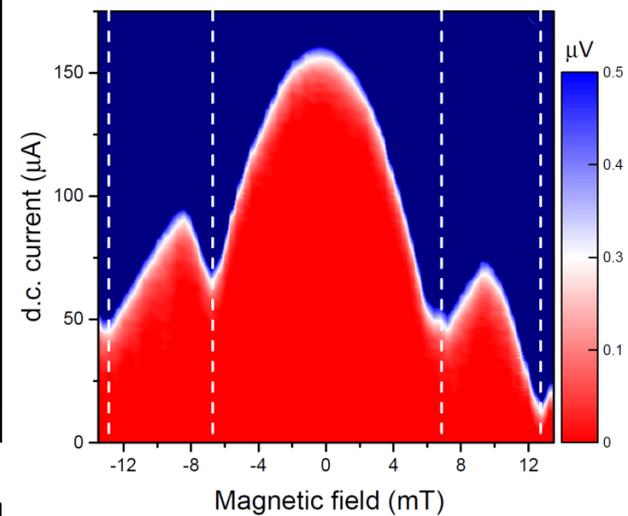
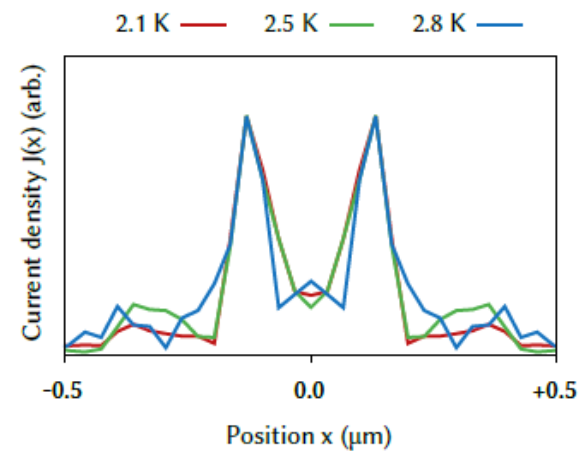

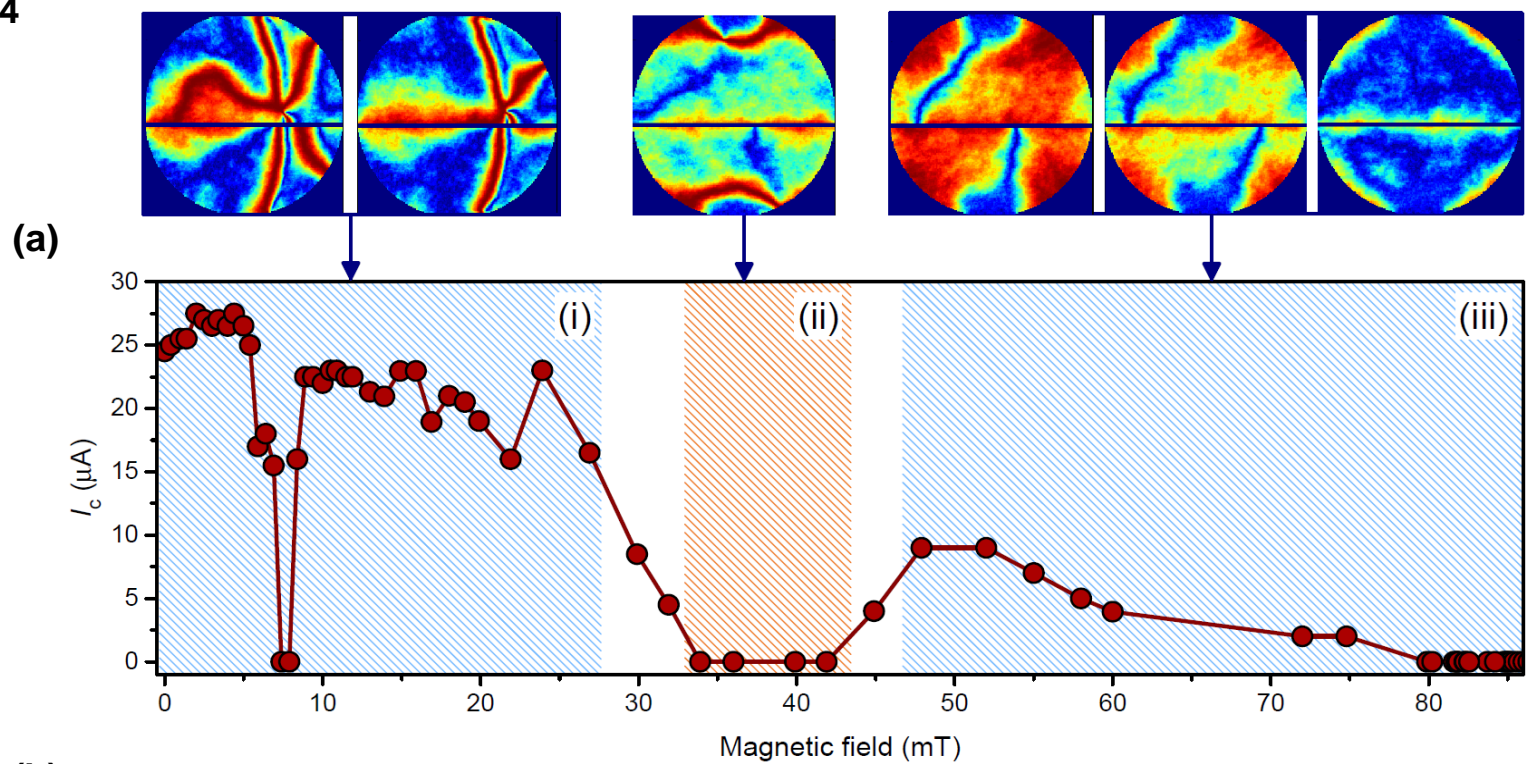

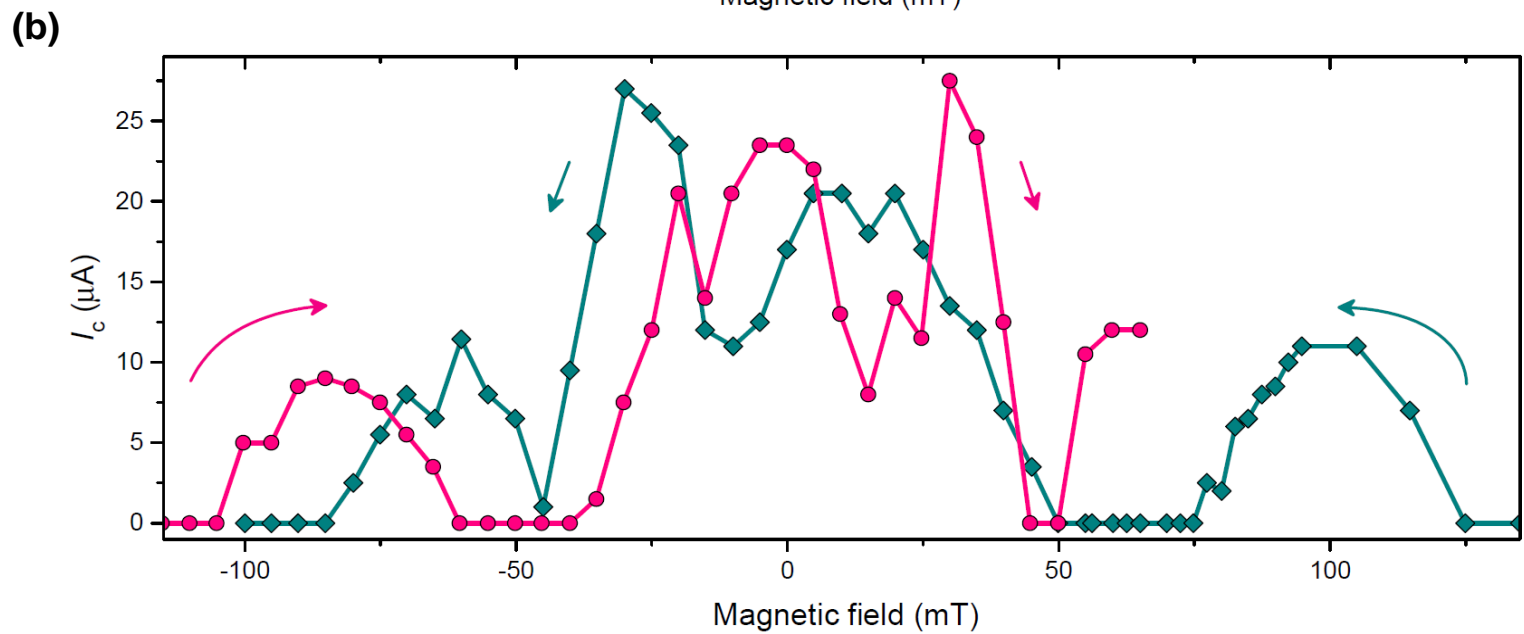

# Supplementary Information

# Controlling supercurrents and their spatial distribution in ferromagnets

Kaveh Lahabi, Morten Amundsen, Jabir Ali Ouassou, Ewout Beukers, Menno Pleijster, Jacob Linder, Paul Alkemade, and Jan Aarts

**The virgin state**

Prior to the measurements presented in the main text, the system was magnetically conditioned by applying an out of plane field of $2.5$ T. This is to reduce the stochastic magnetization introduced by FIB milling of Ni when structuring the junction. The disordered "virgin" state is permanently lost after conditioning, as the system relaxes to its ground state. During this process, the magnetic moments in the Ni layer begin to align with the trench at the centre of the disk. This leads to an increase of MNC in the vicinity of the junction which, in turn, results in an *irreversible* enhancement of triplet correlations at zero field (see Fig. S1).

In contrast to the conditioned sample, $I_c(B_z)$ in the virgin state is suppressed for fields below $5$ mT and the maxima occur at higher fields (as shown in Fig. S2). Note that this offset cannot be attributed to remnant fields from the ferromagnet. The applied field for the SQI measurements is not sufficiently strong to have an appreciable influence on the magnetization (confirmed by SQUID magnetometry and ferromagnetic resonance experiments).

Such unusual interference patterns are the characteristic of systems with multiple parallel $0$ and $\pi$ channels [1,2]. Moreover, it has been proposed that the phase of triplet correlations in a S/F'/F/F'/S junction such as ours, is determined by the relative magnetic orientation of the F' layers on each side of the junction. [3] This could potentially be realized in the virgin state, where the stochastic magnetic orientation of Ni can lead to random formation of multiple $0$ and $\pi$ segments across the junction. Furthermore, these interference patterns are characterized by irregular discontinuities. This can be attributed to the arbitrary arrangement of $0$ and $\pi$ segments. Remarkably, we find these features to disappear altogether after conditioning the sample: $I_c$ is persistently higher at zero field, and the interference patterns have become highly regular and reproducible. Furthermore, the observed dependency on conditioning was entirely absent in junctions without the Ni layer as we found no discernable changes in the interference pattern or the magnitude of $I_c(B=0)$.

**Micromagnetic simulations and critical current measurements with in-plane magnetic field**

The leads used for transport measurements are included for simulations with in-plane magnetic fields (Fig. S3). While the leads do not disrupt the stability of the magnetic vortex, it leads to the emergence of a domain wall in Ni. In the absence of in-plane field, the overall supercurrent distribution across the junction remains unaffected by the leads, as the vortex core continues to suppress the MNC (resulting in two transport channels). However, the influence of the leads on shape anisotropy becomes relevant when magnetizing the sample with $B \parallel y$. This allowed for an accurate estimate of the MNC and the corresponding variations in $I_c$ during a field sweep. As shown in Fig. S4 the simulated field for vortex entry (exit) corresponds to a distinct enhancement (suppression) of the measured $I_c$, marked by **i** (**ii**).

The observed hysteresis is different from the usual hysteresis observed in SFS junctions, where the self-field of the ferromagnet(s) results in a shift in the measured Fraunhofer pattern [4,5,6]. Rather, it is a distinctive characteristic of triplet generation being realized by the magnetic misalignment of multiple layers [7]. The resulting $I_c(B_y)$ pattern is complex, and yet highly reproducible, where individual features are mirrored (and not just shifted) with respect to the direction of field sweep. The most notable difference here is arguably the relatively large field range where $I_c$ is zero and the pronounced reentrant superconductivity that follows.

**Numerical simulations of the critical current**

To calculate the critical current we use the quasiclassical approximation in the diffusive limit, which yields the Usadel equation [8]

$$D\nabla \hat{g} \nabla \hat{g} + i[\varepsilon \hat{\rho}_3 + \hat{\boldsymbol{\sigma}} \cdot \boldsymbol{h}, \hat{g}] = 0$$

where $D$ is the diffusion constant and $\varepsilon$ is the quasiparticle energy. The magnetization texture from the micromagnetic simulations are represented as an exchange field $\boldsymbol{h} = \boldsymbol{h}(\boldsymbol{r})$. Furthermore we have defined $\hat{\boldsymbol{\sigma}} = \text{diag}(\boldsymbol{\sigma}, \boldsymbol{\sigma}^*)$, where $\boldsymbol{\sigma}$ is a vector of Pauli matrices, and $\hat{\rho}_3 = \text{diag}(1,1,-1,-1)$. From $\hat{g} = \hat{g}(\boldsymbol{r}, \varepsilon)$, the $4 \times 4$ retarded Green function matrix in Nambu $\otimes$ spin space, the equilibrium current density may be computed as

$$\boldsymbol{J} = \frac{N_0 e D}{2} \int d\varepsilon \, \text{Re} \, \text{Tr}\{\hat{\rho}_3 \hat{g} \nabla \hat{g}\} \tanh \frac{\beta \varepsilon}{2}$$

where $N_0$ is the density of states at the Fermi level, and $\beta = 1/k_B T$. We neglect the inverse proximity effect, and assume that the superconductors on each side of the trench are large enough to be approximated as bulk. In the calculations, we have used that the critical current is approximately found for a phase difference between the superconductors of $\Delta \phi = \frac{\pi}{2}$. For simplicity, we use transparent boundary conditions between the Ni and the Co layer, whereas we use the low-transparency Kupriyanov-Lukichev boundary conditions [9] at the Ni-Nb interface.

In the modeling of the geometry, we have assumed an effective superconducting coherence length of $\xi = 10$ nm, so that the radius of the circular disk becomes $R = 50\xi$. In the direction crossing the trench, the model has been truncated to a width of $W = 40\xi$ to reduce the model size. This has been done under the assumption that any contribution to the current from the removed regions are negligible due to the vast distance to the opposite superconductor. The thickness of the Ni and the Co layers have been set to $\xi$ and $6\xi$, respectively, and the width of the trench is $2\xi$.

The spatial distribution of the magnetization in both the Ni and the Co layer are accurately mapped onto the 3D mesh via the exchange field $\boldsymbol{h}$, and we have used an amplitude $|\boldsymbol{h}| = 30\,\Delta \simeq 46$ meV. While this is significantly lower than typical exchange fields in Co, it is still sufficient to quench the contribution to the current density from singlet Cooper pairs. To verify this, we make use of the fact that the supercurrent density generated by the singlet Cooper pairs, $\boldsymbol{J}^{(s)}$, and the triplet Cooper pairs, $\boldsymbol{J}^{(t)}$, contribute additively; $\boldsymbol{J} = \boldsymbol{J}^{(s)} + \boldsymbol{J}^{(t)}$. The two components are shown in Fig. S5, where it can be seen that the current density generated by the singlet Cooper pairs quickly decays away from the superconductors. The triplet current density on the other hand, remains appreciable different from zero in a significantly larger region, indicating that it is the primary means of transport. The results will therefore be qualitatively the same as with a more realistic strength of the exchange field. The advantage of using the reduced value is that the current densities become larger, which in turn make the numerical calculations less resource intensive.

In the finite element analysis there has been used 27-node hexagonal volume elements, and the Green function is interpolated within each element by means of second order Lagrange polynomials. This means that the current density within each element is interpolated by linear polynomials. To ensure that the spatial distribution of the current density is accurately resolved, we use a refined mesh in a region surrounding the trench, as is shown in Fig. 2a in the main text. For more details regarding the finite element analysis of three-dimensional superconducting heterostructures, please consult Ref.[10].

**Fourier analysis of the field-dependent supercurrent profiles**

As shown by Dynes and Fulton [11], the supercurrent density profile $J(x)$ can be determined from the superconducting interference pattern $I_c(B)$ using a Fourier transform:

$$J(x) \sim \int_{-\infty}^{+\infty} dB\, I_c(B)\, e^{2\pi i L B x/\Phi_0}$$

Here, the coordinate system is defined such that the magnetic field $B$ is applied along the $z$-axis, the critical current $I_c$ is measured along the $y$-axis, and the current distribution $J(x)$ can then be determined along the $x$-axis. The equation also depends on the effective length $L$ of the junction and the flux quantum $\Phi_0 = h/2e$. Note that $I_c(B)$ is the *signed* critical current, where the sign is determined from the experimentally measured $|I_c(B)|$ by assuming that it consists of alternating positive and negative lobes, as described in more detail in Ref.[11]. This procedure is justified when the interference pattern consists of well-defined maxima separated by deep minima, as is the case for our measurements.

The original method by Dynes and Fulton was derived for a rectangular junction where the dimensions of each superconductor are much larger than the London penetration depth $\lambda$. In that case, the effective junction length $L = 2\lambda + d$, where $d$ is the thickness of the barrier between the superconducting leads. In our case, however, the effective junction length is limited by the geometry and not the penetration depth, which follows from the fact that the measured oscillation period is temperature-independent. Furthermore, we have a planar cylindrical junction, so the effective length is not uniform. However, the Fourier analysis can still provide a qualitatively correct picture, especially near the center of the junction where the junction length is nearly constant. It is however difficult to estimate the appropriate junction length *a priori*. We therefore performed the Fourier analysis without making any assumptions regarding the value of $L$, and instead assumed that the points where we obtained $J(x) \to 0$ likely corresponded to the junction ends $x \approx \pm R$, where $R$ is the cylinder radius. From this, we obtained an estimate $L \approx 180$ nm for the effective junction length. This value agrees with the effective area of the junction as comes out of the period of the $I_c(B_z)$ oscillations. If we were to take the sharp drop in the current density profile as the sample edge, $L$ would become less than 100 nm, leading to an unphysically low effective area as compared to the actual disk radius of $R = 0.5\ \mu$m.

The SQI experiments are carried out by measuring the voltage as a function of current for a given applied magnetic field, i.e. $V(I, B)$. The critical current $|I_c(B)|$, used for the Fourier analysis, is obtained by extracting a contour for a small but finite voltage threshold $V(I_c, B) > 0.3\ \mu$V. Experimentally we find this criterion to be optimal for reducing noise effects that distort the shape of $I_c(B)$. The result is then adjusted to the $y$-axis so that $|I_c(B)| = 0$ at the nodes between the lobes of the interference pattern. This is to account for the artificial offset introduced by the 0.3 μV threshold voltage. We then recover the complex critical current $I_c(B)$, by switching the sign of every other lobe of the measured $|I_c(B)|$. The original $|I_c(B)|$ and the signed $I_c(B)$ curves are shown side-by-side in Fig. S6.

Note that the measured $I_c(B)$ may slightly deviate from a perfectly symmetric pattern, and yield a *complex* supercurrent distribution $J(x)$ after Fourier transformation. This apparent asymmetry however is predominantly caused by experimental noise. We therefore discard the complex phase $J(x)$ to approximate the supercurrent distribution profile by $|J(x)|$, shown in Fig. 3f of the main text.

## Captions

**Supplementary Fig. S1| Comparing basic transport properties before and after conditioning the magnetic state. a,** Resistance as a function of temperature before (pink) and after (blue) conditioning the sample. In each set, two distinct transitions are observed. At $T = 5.5$ K, the Nb electrodes become superconducting, while the junction is still in the normal state ($R_N \approx 240$ mΩ). Upon cooling further, resistance undergoes a second transition as the junction begins to proximize by triplet correlations — eventually reaching zero resistance. While the superconducting electrodes are unaffected by conditioning, we observe substantial enhancement of superconductivity in the junction. For clarity, the inset shows the transition for both states on a logarithmic scale. The influence of conditioning is also reflected in the $I - V$ curves taken before **(b)** and after **(c)** conditioning.

**Supplementary Fig. S2| SQI patterns for the virgin magnetic state. a,** Disordered magnetic state of Ni before conditioning (schematic). The stochastic magnetic orientation of Ni on each side of the trench can lead to the formation of multiple 0 and $\pi$ segments across the junction. **b,** SQI pattern of the virgin sample measured while sweeping the field $B_z$: $-30 \rightarrow 30$ mT in steps of 0.3 mT. On average, the supercurrent is suppressed for small fields (below 5 mT) in *both* field directions. The SQI pattern is characterized by random discontinuities. These irregularities are shown more clearly in **c,** where individual $I - V$ curves are plotted for $B_z$: $28 \rightarrow 0$ mT. The curves are given an offset to represent the field they were measured at. All measurements are taken at $T = 2.1$ K.

**Supplementary Fig. S3| Micromagnetic simulations with an in-plane field**. Top views of the magnetic states of Co and Ni layers with the addition of the leads for transport measurements, obtained from OOMMF simulations. Individual components of the magnetization vector ***m*** are plotted separately for clarity. The pixel colour scheme, red-white-blue, scales with the magnitude of $m$ for each component. The red and blue pixels represent positive and negative values respectively. Out-of-plane magnetization ($m_z$) is generally suppressed, except at the vortex core where both layers have a highly localized out-of-plane component. In the actual device, the trench that forms the junction is slightly off-centred. This feature is accounted for in the simulations by placing the gap in Ni 40nm away from the centre of the disk. Including the leads in the simulated design modifies the shape anisotropy of the system. This becomes particularly relevant for simulations where the field is applied along $y$.

**Supplementary Fig. S4 | Simulated magnetic non-collinearity and the measured critical current variation with in-plane magnetic field. a,** Snapshots of simulated magnetic non-collinearity at different stages of magnetization reversal (left) and the experimental values of critical current measured for $B_y$: $200$ mT $\rightarrow -200$ mT. Taking steps of 5 mT, simulations show the vortex enters the system at $-20$ mT, resulting in an enhancement of MNC. At $-45$ mT the vortex exits the system. This leads to substantial increase in stray fields, which magnetize the Ni layer antiparallel to Co, hence reducing the MNC. Increasing the field further, the Ni layer begins to magnetize for a second time (along $-y$). This increases the MNC, reaching a maximum at $-60$ mT. Beyond this field however, both Co and Ni magnetization begin to align, and MNC declines (as shown for $B_y = -85$ mT). The $I_c$ values for the corresponding fields are marked with red circles. Stages **i** and **ii** represent vortex entry and exit, while **iii** corresponds to the second magnetization reversal of Ni. **b,** The same pattern $I_c$ (i.e. stages **i** − **iii**) is observed when sweeping the field in opposite direction,

$B_y$: $-200$ mT $\rightarrow$ 60 mT (pink). Finally at 60 mT, the field is reversed back to zero (blue). This initially reduces the MNC by allowing local stray fields to magnetize Ni antiparallel to Co, and $I_c$ declines to zero. At 20 mT however, the vortex enters the Co again, and the stray fields are effectively suppressed. This restores the MNC and we observe a sharp increase in $I_c$.

**Supplementary Fig. S5 | Simulated supercurrent density contributions from singlet and triplet Cooper pairs. a,** Magnitude of the current density generated by singlet Cooper pairs, $J^{(s)} = |\boldsymbol{J}^{(s)}|$, which is greatly suppressed except for in the immediate vicinity of the superconductors. **b,** Magnitude of the current density generated by triplet Cooper pairs, $J^{(t)} = |\boldsymbol{J}^{(t)}|$. For clarity, currents lower than $10^{-7} J_0$ have been removed, which explains why no singlet current is observed in the trench. It is noted that while the total current $\boldsymbol{J} = \boldsymbol{J}^{(s)} + \boldsymbol{J}^{(t)}$ is conserved, $\boldsymbol{J}^{(s)}$ and $\boldsymbol{J}^{(t)}$ are generally not. This is due to the magnetization, which causes oscillations between the singlet and triplet states.

**Supplementary Fig. S6 | Recovering the complex critical current**. **a,** The (unsigned) $|I_c(B)|$ pattern extracted from $I-V$ measurements. **b,** The signed $I_c(B)$ interference pattern reconstructed by flipping the signs of alternate lobes as in Ref.[11]. The data were tanek at 2.1 K.

**Supplementary Fig. S1**

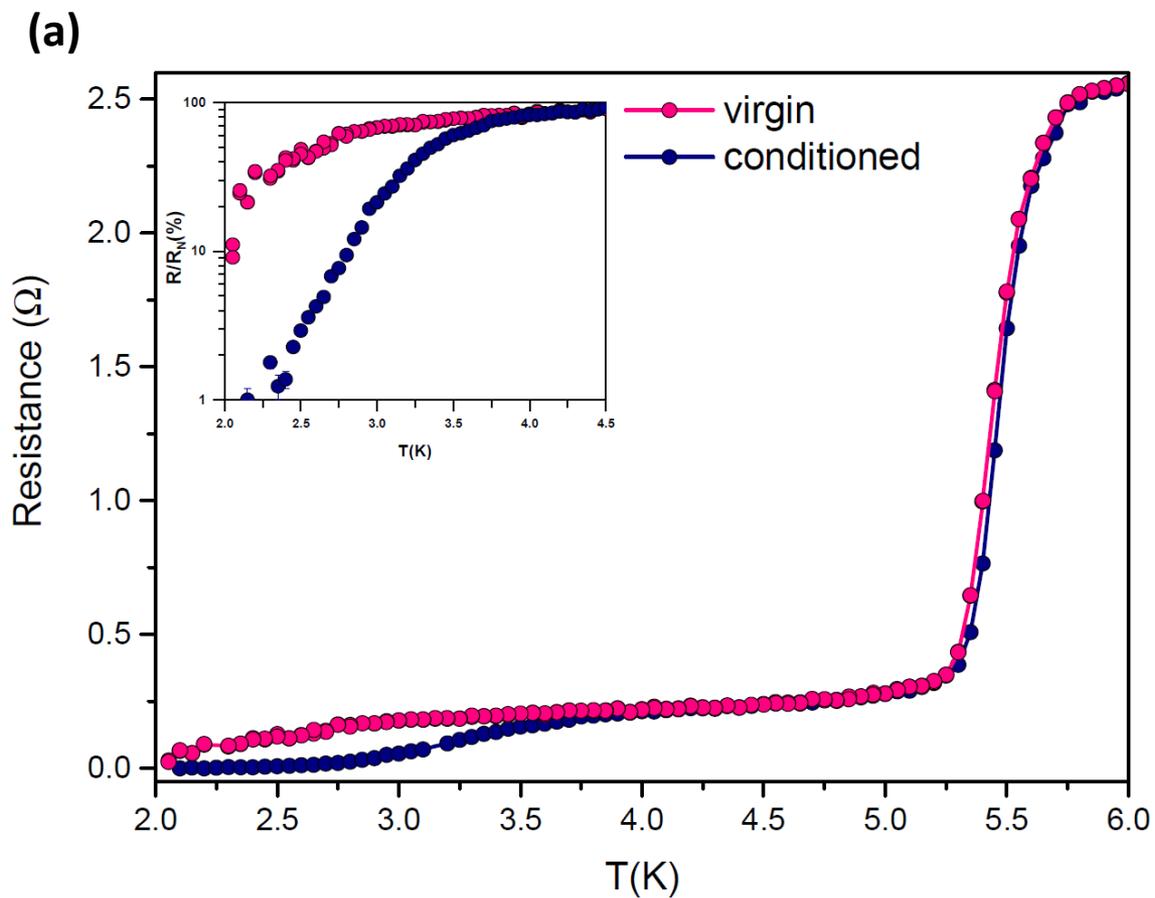
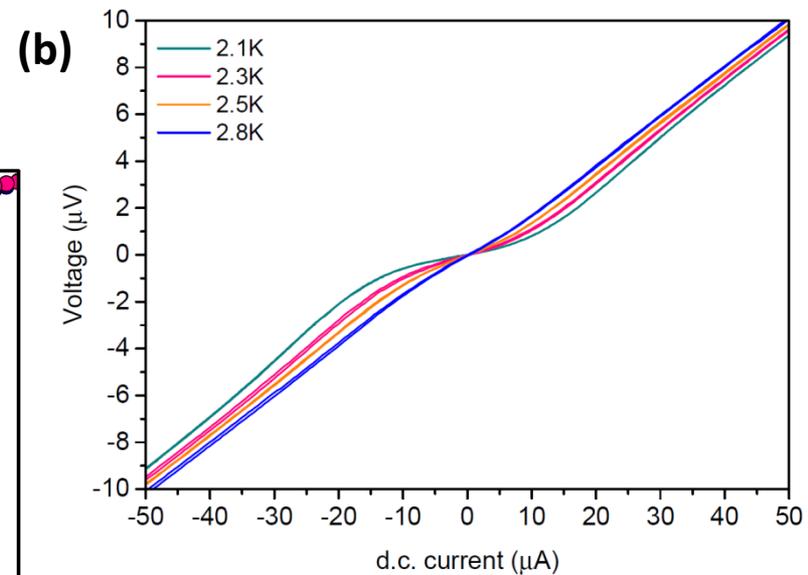
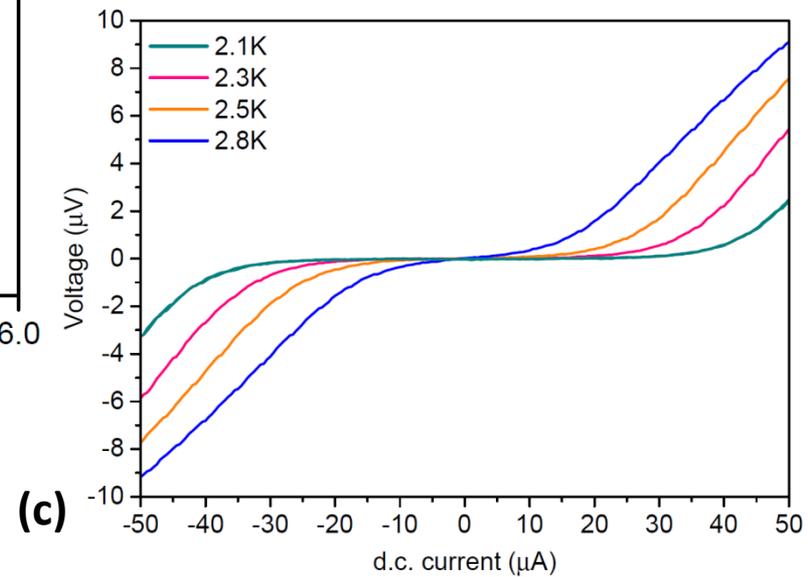

**Supplementary Fig. S2**

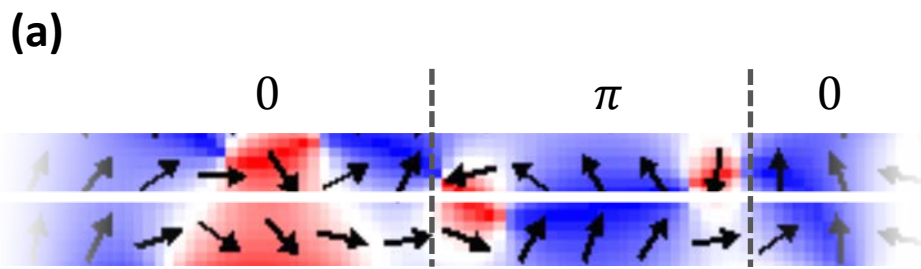

(a)

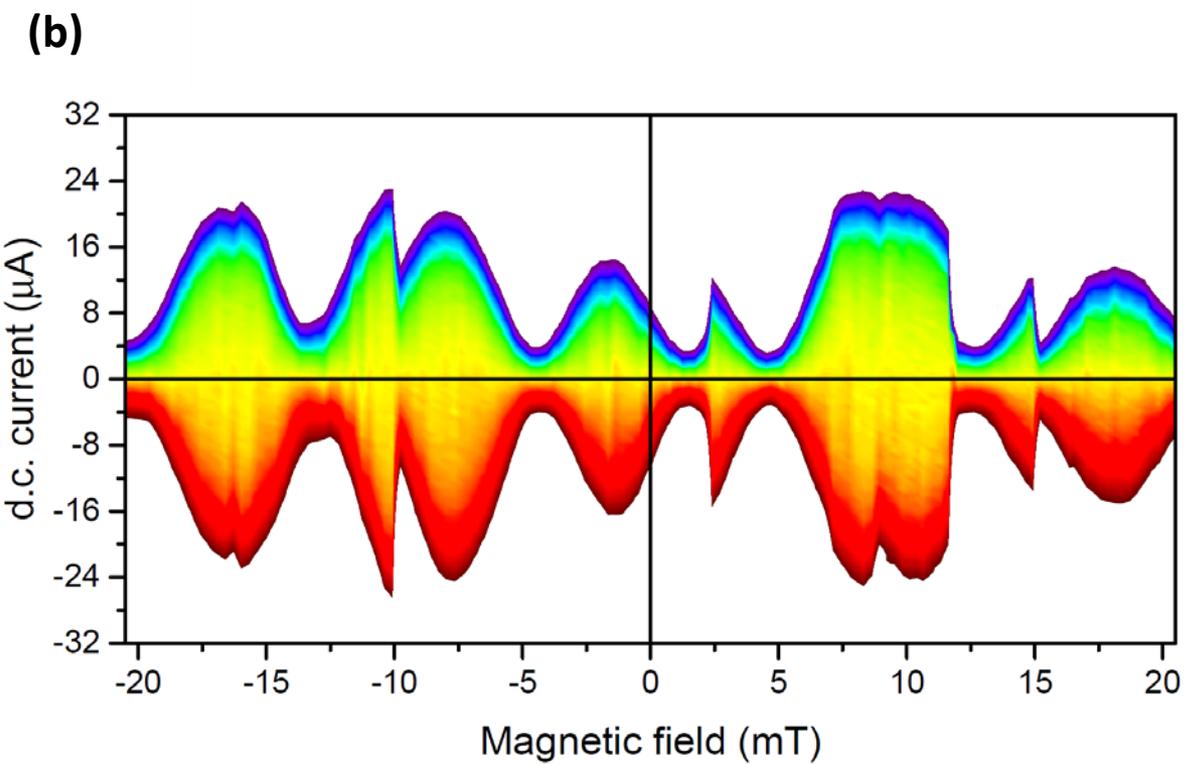

(b)

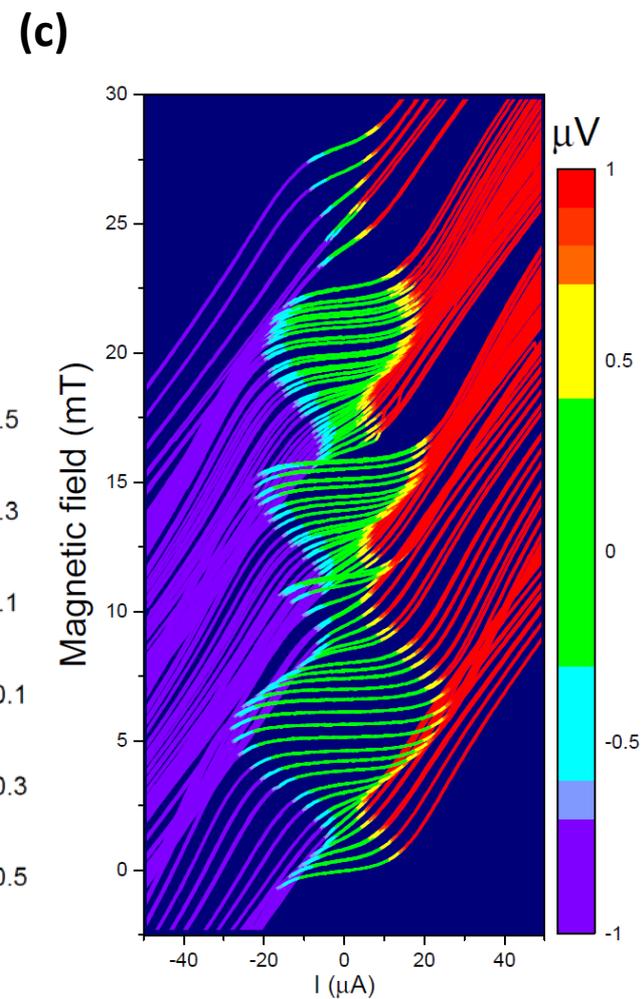

(c)

**Supplementary Fig. S3**

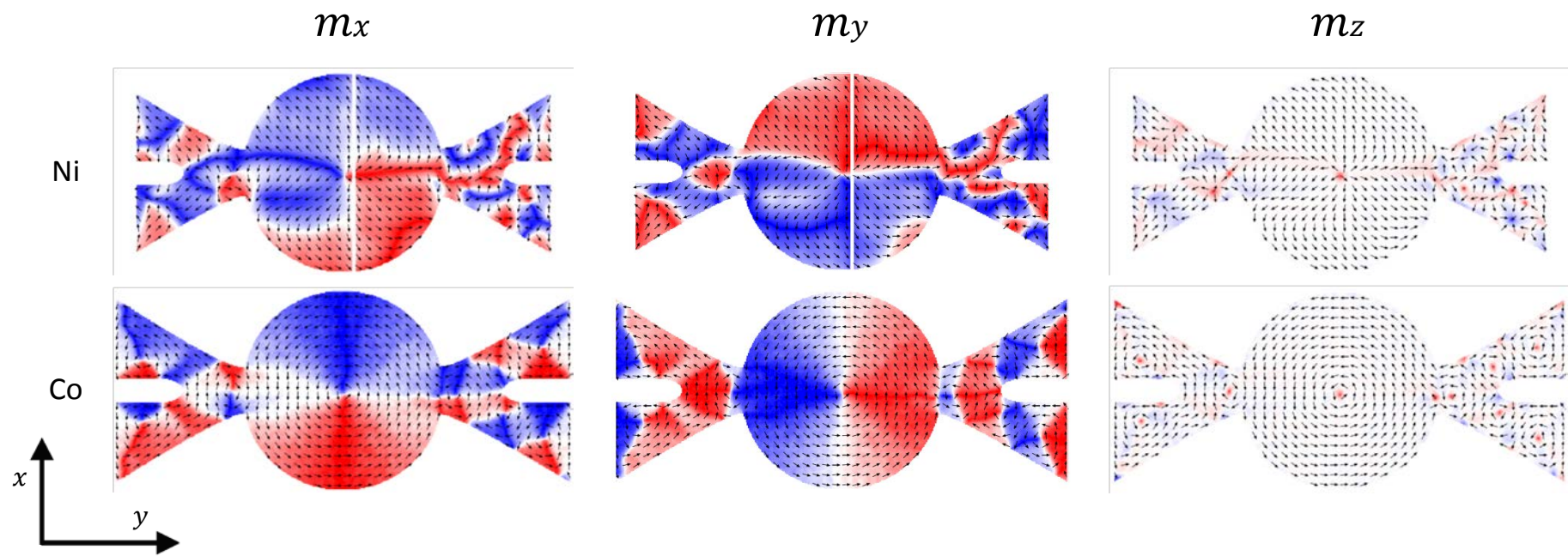

**Supplementary Fig. S4**

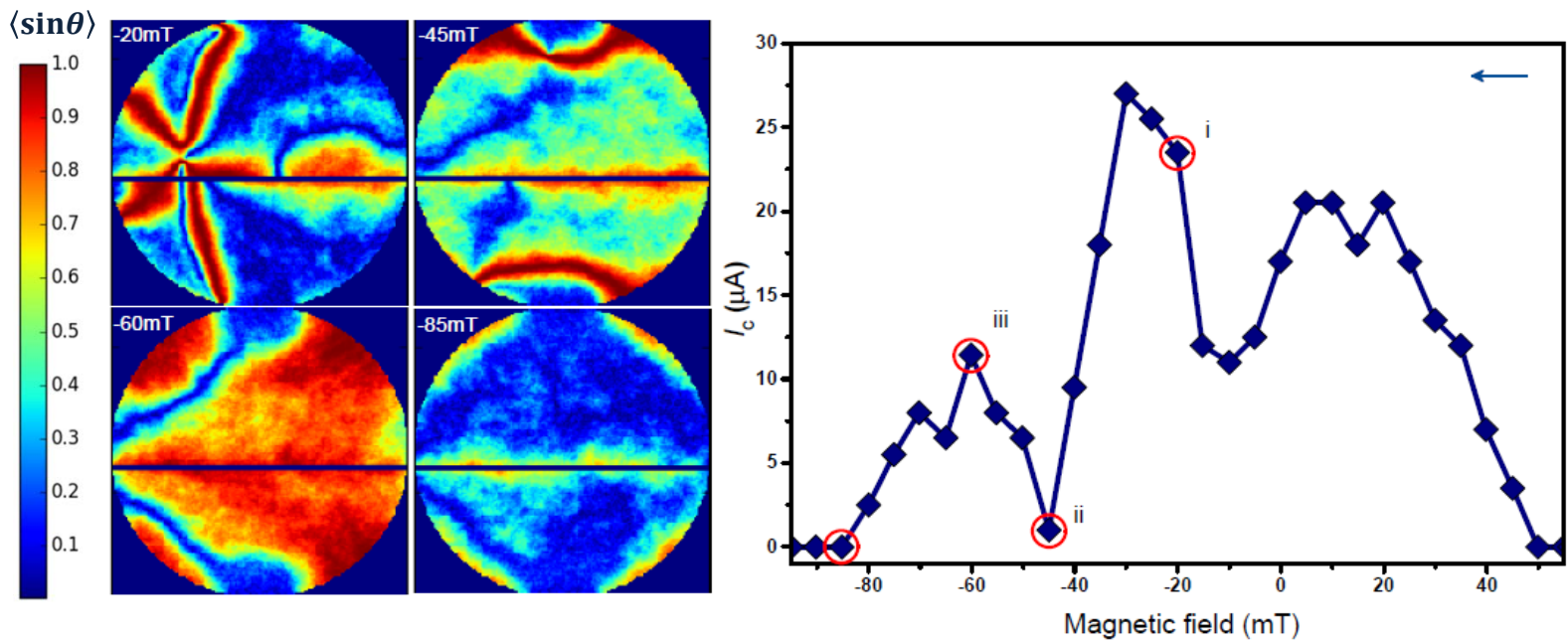

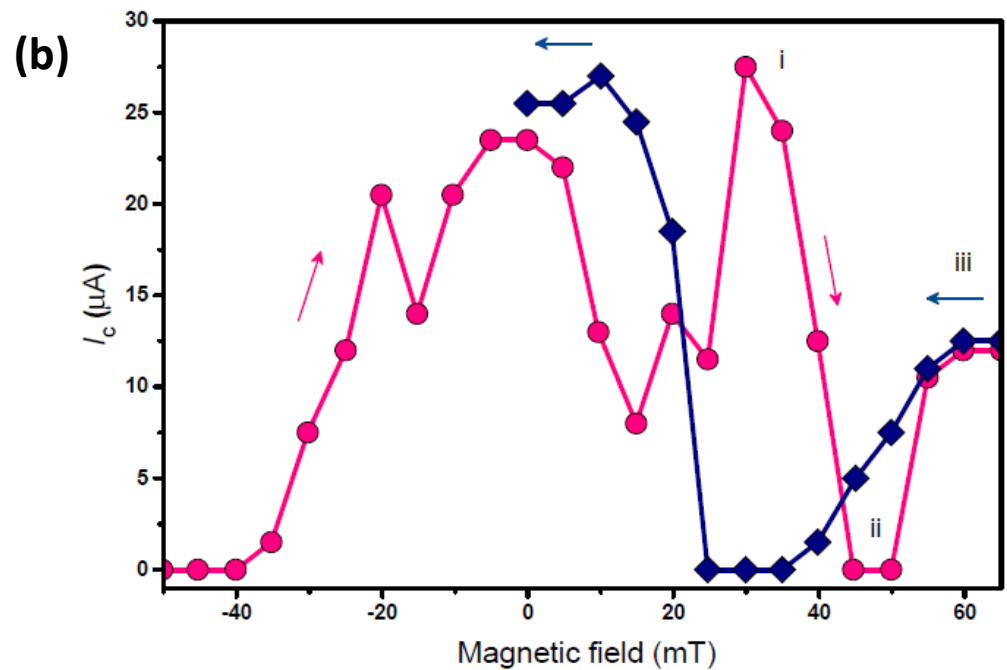

**Supplementary Fig. S5**

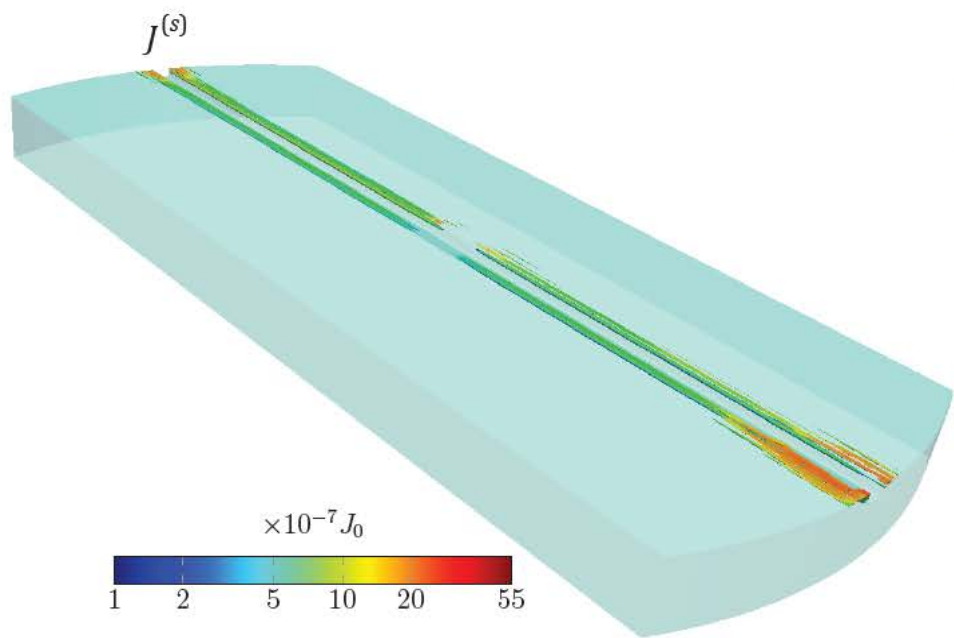
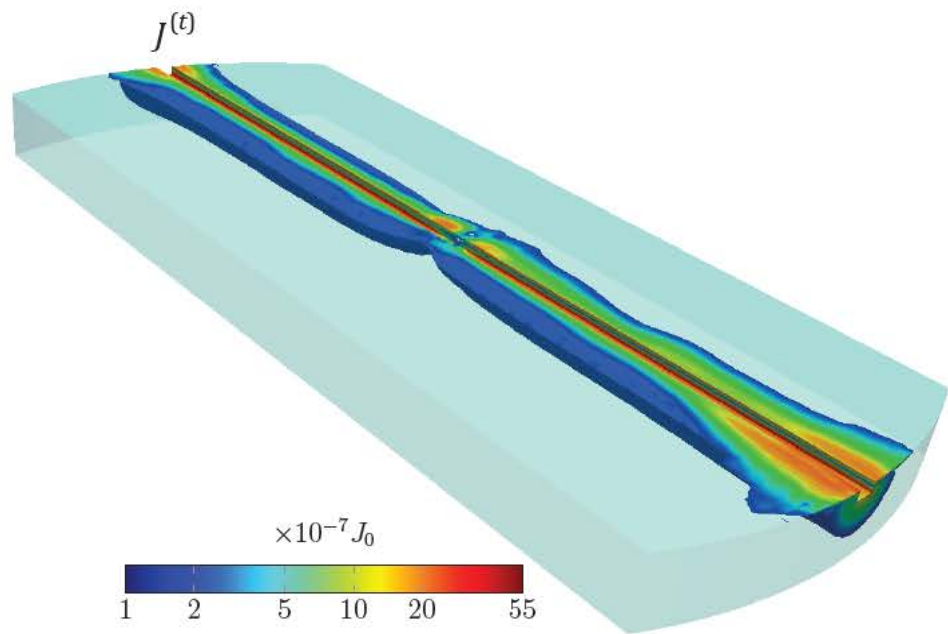

(a) (b)

**Supplementary Fig. S6**

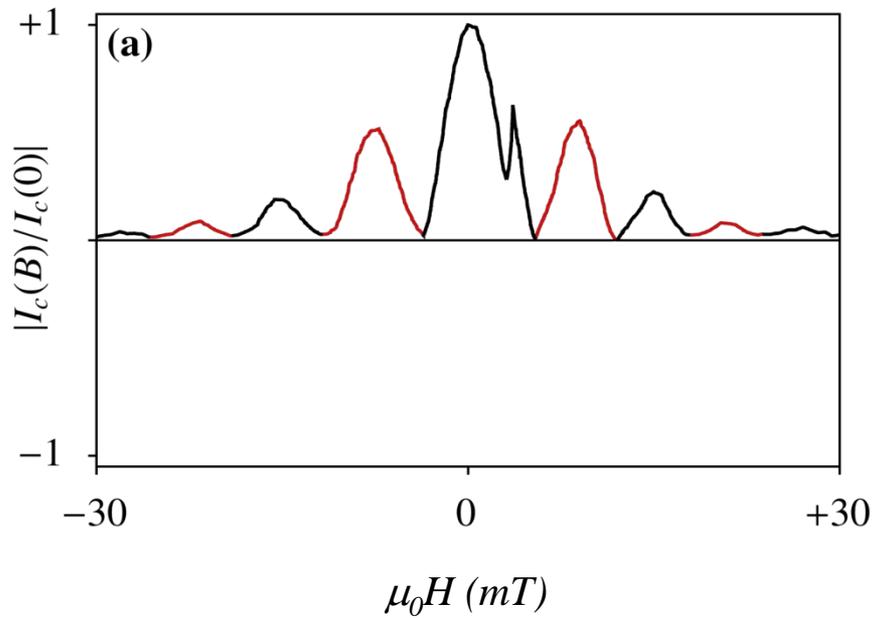 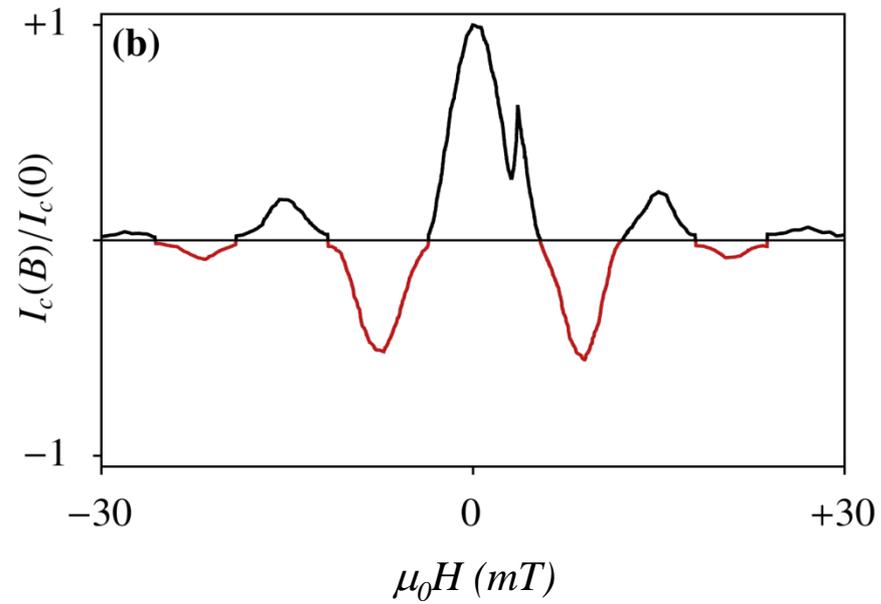